\documentclass[twocolumn,prl]{revtex4}

\usepackage{graphicx}
\begin{document}

\title{The Electronic and Superconducting Properties of Oxygen-Ordered {MgB$_{2}$} compounds of the form
{Mg$_{2}$B$_{3}$O$_{x}$}}

\author{Juan C. Idrobo}
\affiliation{Department of Physics, University of
California-Davis, Davis, One Shields Ave, Davis CA, 95616, USA}
\email{jidrobo@lbl.gov}

\author{Serdar \"{O}\u{g}\"{u}t}
\affiliation{Department of Physics, University of Illinois at
Chicago, 845 West Taylor Street (M/C 273), Chicago, IL 60607}

\author{Taner Yildirim}
\affiliation{NIST Center for Neutron Research, National Institute
of Standards and Technology, Gaithersburg, Maryland 20899 USA}

\author{Robert F. Klie}
\affiliation{Brookhaven National Laboratory, Upton, New York
11973}

\author{Nigel D. Browning}
\affiliation{Dept. of Chemical Engineering and Material Science,
University of California-Davis, One Shields Ave, Davis CA, 95616,
USA}
\affiliation{National Center for Electron Microscopy, Lawrence
Berkeley National Laboratory, Berkeley, CA 94720 USA}

\date{\today}

\begin{abstract}
Possible candidates for the {Mg$_{2}$B$_{3}$O$_{x}$}
nanostructures observed in bulk  of polycrystalline {MgB$_{2}$}
\cite{klie2002-01} have been studied using a combination of
Z-contrast imaging, electron energy loss spectroscopy (EELS) and
first-principles calculations. The electronic structures, phonon
modes, and electron phonon coupling parameters are calculated for
two oxygen-ordered {MgB$_{2}$} compounds of composition
{Mg$_{2}$B$_{3}$O} and {Mg$_{2}$B$_{3}$O$_{2}$}, and compared with
those of {MgB$_{2}$}. We find that the density of states for both
{Mg$_{2}$B$_{3}$O$_{x}$} structures show very good agreement with
EELS, indicating that they are excellent candidates to explain the
observed coherent oxygen precipitates. Incorporation of oxygen
reduces the transition temperature and gives calculated
\textit{T$_{C}$} values of 18.3 K and 1.6 K for {Mg$_{2}$B$_{3}$O}
and {Mg$_{2}$B$_{3}$O$_{2}$}, respectively.
\end{abstract}

\pacs{} \maketitle

The discovery of superconductivity in {MgB$_{2}$} with a
transition temperature of 40 K \cite{naga2001-01} has attracted
the attention of the scientific community for two main reasons:
the technological applications of this material and the new
insights that such a simple structure could bring to
superconductivity theory. Efforts to improve the superconductivity
properties of {MgB$_{2}$} have included doping with elements such
as Y \cite{wang2002-01}, Zr \cite{feng2001-01}, C
\cite{take2001-01}, Al \cite{slusky2001-01}, Cu and Ag
\cite{solta2002-01}, but the results varied for different groups.
A simple explanation for this behavior is that the transport
properties of {MgB$_{2}$} have a strong dependence on the sample
preparation conditions, which makes it hard to obtain unambiguous
results even for undoped samples \cite{yan2003-01}. In fact, there
have been reports from different groups showing that {MgB$_{2}$}
samples are not a single phase material, but rather a rich
collection of different phases including {MgB$_{2}$}, MgO,
{B$_{y}$O$_{x}$}, {Mg$_{x}$B$_{y}$}, {Mg$_{x}$B$_{y}$O$_{z}$}
\cite{klie2002-01,sharma2002-01,klie2001-01,liao2003-01}. Oxygen,
unlike some other elements, is present in {MgB$_{2}$} as an
unintentional impurity due to its high reactivity with
{MgB$_{2}$}. Typically, oxygen rich precipitates form in the
bulk of {MgB$_{2}$} where they contribute to the overall transport
properties through flux pinning \cite{liao2003-01}. Moreover,
oxygen can segregate at grain boundaries of {MgB$_{2}$} where,
unlike the high$-T_c$ materials, it
contributes significantly to flux pinning, increasing the overall
{\em{J$_{C}$}} as the grain size decreases \cite{larba2001-01}.
{MgB$_{2}$} has already been used in making wires
\cite{canfi2001-01}, tapes \cite{grasso2001-01}, and Josephson
junctions \cite{mija2002-01}, and in order to improve the
properties of such devices, the effect that oxygen has on the
properties of {MgB$_{2}$} needs to be well understood.
Surprisingly, the importance of oxygen in the development of
{MgB$_{2}$} devices has been overlooked, and detailed
first-principles studies of oxygen ordering and segregation have
not yet been performed.

\begin{figure}
\includegraphics[scale=0.6]{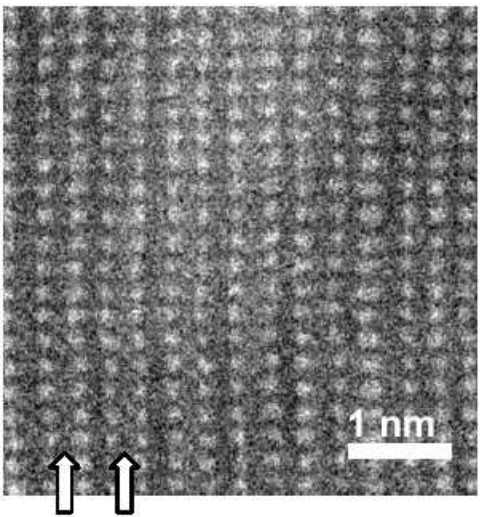}
\caption{Z-contrast image of a coherent oxygen precipitate in the
bulk of {MgB$_{2}$} [010]. The bright spots are Mg atoms. The
increase of contrast every second column is due to presence of
oxygen every second boron plane. White arrows highlight the atomic
columns where oxygen has precipitated.}
\end{figure}

It was previously reported that oxygen impurities can increase the
upper critical fields and critical current but lower
{\em{T$_{C}$}} \cite{eom2001-01}. Liao {\em{et. al.}} found that
through a slow cooling process, the oxygen in bulk {MgB$_{2}$} can
segregate to form nanometer-sized coherent Mg(B,O) in the
{MgB$_{2}$} matrix acting as effective flux pinning centers
without decreasing {\em{T$_{C}$}} \cite{liao2003-01}. Klie
{\em{et. al.}} previously reported oxygen precipitates in bulk
{MgB$_{2}$}, where oxygen was incorporated every second boron
plane of the {MgB$_{2}$} structure, found by atomic resolution
electron energy loss spectroscopy (EELS) studies using a scanning
transmission electron microscope \cite{klie2002-01}. Figure 1
shows an atomic resolution Z-contrast image of one of these oxygen
precipitates taken from bulk {MgB$_{2}$} along the [100]
orientation. The increase of contrast every second column in Fig.
1 is due to the presence of oxygen \cite{browning1993-01}. The
size of these precipitates was reported to be about 20-100 nm
\cite{klie2002-01}.

Motivated by these experimental observations, in this work we
discuss the effects of oxygen-ordering, with different
stoichiometries, on the electronic structure and superconductivity
properties of {MgB$_{2}$} via first-principles calculations. As
possible candidates for the atomic structures of coherent oxygen
precipitates in bulk {MgB$_{2}$}, we propose two different
structures of composition, {Mg$_{2}$B$_{3}$O} and
{Mg$_{2}$B$_{3}$O$_{2}$}. We find that incorporation of oxygen in
the form of {Mg$_{2}$B$_{3}$O$_{x}$} structures decreases the DOS
at the Fermi energy and the electron phonon coupling compared with
{MgB$_{2}$}. From total energy calculations we find that
{Mg$_{2}$B$_{3}$O} is less stable compared to
{Mg$_{2}$B$_{3}$O$_{2}$} but has a higher {\em{T$_{C}$}}. It is
important to notice that our calculations show that oxygen
impurities are likely to be present everywhere in {MgB$_{2}$}
structures. Therefore, controlling the density and type of these
impurities will finally be the limiting factor in improving the
transport properties of {MgB$_{2}$}.

Our atomic and electronic structure calculations were performed
using the ab initio pseudopotential plane wave method
\cite{kresse1996-01} within the generalized gradient approximation
(GGA) \cite{perdew1992-01}, and local density approximation (LDA)
for comparison. In order to compare
the changes due to oxygen in {MgB$_{2}$}, we doubled the unit cell
of {MgB$_{2}$} in the c-direction. Brillouin-zone integrations
were carried out with 15$\times$15$\times$6 Monkhorst-Pack grids
for {Mg$_{2}$B$_{4}$} and {Mg$_{2}$B$_{3}$O$_{x}$}. An energy
cutoff of 400 eV was used for all the calculations. The lattice
volumes, shapes, and all atoms were allowed to relax for the new
structures using a force criterion of $10^{-2}$ eV/\AA. The
zone-center phonons and electron phonon couplings were calculated
using density functional perturbation theory \cite{baroni2001-01}.

\begin{figure}
\includegraphics[scale=0.2]{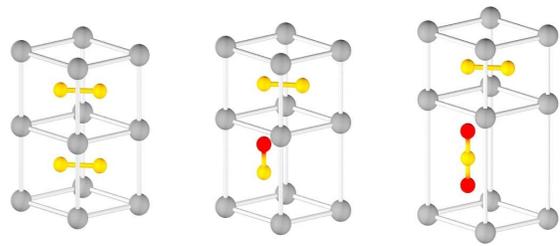}
\caption{(color). From left to right: schematics of
{Mg$_{2}$B$_{4}$}, {Mg$_{2}$B$_{3}$O} and {Mg$_{2}$B$_{3}$O$_{2}$}
crystal structures.  The lattice in the c-direction stretches for
{Mg$_{2}$B$_{3}$O$_{2}$} structures with respect to
{Mg$_{2}$B$_{4}$}, where the addition of oxygen increases the
distance between the Mg planes where oxygen was added. Mg, B, and
O atoms are shown in gray, yellow, and red respectively.}
\end{figure}

\begin{table*}
\caption{\label{table1}Summary of properties for {MgB$_{2}$},
{Mg$_{2}$B$_{4}$}, {Mg$_{2}$B$_{3}$O} and {Mg$_{2}$B$_{3}$O$_{2}$}
crystal structures. The {Mg$_{2}$B$_{3}$O$_{x}$} structures belong
to the space group P3m1 ({\em{C}$_{3v}^{1}$}, No. 156).  Lattice
parameters are in \AA, $N(E_{F})$ in states/eV, frequencies
($\omega$) in meV. The parenthesis next to the frequencies
indicates degeneracy of those modes. {\em{T$_{C}$}} is given in K.
The lattice parameters are shown for GGA (LDA) approximations. The
phonon modes, electron phonon coupling, and the critical
temperatures were all calculated using the values from the GGA
calculations.}
\begin{ruledtabular}
\begin{tabular}{cccccccc}
\multicolumn{2}{c}{MgB$_{2}$}&\multicolumn{2}{c}{Mg$_{2}$B$_{4}$}&\multicolumn{2}{c}{Mg$_{2}$B$_{3}$O}
&\multicolumn{2}{c}{Mg$_{2}$B$_{3}$O$_{2}$}\\\hline

\multicolumn{2}{c}{a=3.086\footnote{Taken from Ref.
\cite{naga2001-01}}}&\multicolumn{2}{c}{a=3.046(3.086)}&\multicolumn{2}{c}{a=3.067(3.016)}&\multicolumn{2}{c}{a=3.143(3.016)}\\

\multicolumn{2}{c}{c=3.524\footnotemark[1]}&\multicolumn{2}{c}{c=7.054(7.048)}&\multicolumn{2}{c}{c=7.701(7.545)}&\multicolumn{2}{c}{c=8.633(8.231)}\\

\multicolumn{2}{c}{$N(E_{F})$=0.69\footnote {\mbox{Values for
{\em{N(E$_{F}$)}}, $\omega$, $\lambda$ and {\em{T$_{C}$}} (with
{$\mu^{*} = 0.15$}) are taken from Ref.
\cite{yildi2001-01}}}}&\multicolumn{2}{c}{{\em{N(E$_{F}$)}}=1.35(1.38)}
&\multicolumn{2}{c}{{\em{N(E$_{F}$)}}=0.85(0.87)}
&\multicolumn{2}{c}{{\em{N(E$_{F}$)}}=1.05(1.99)}

\\\hline

$\omega_{p}$\footnotemark[2]&$\lambda_{p}$\footnotemark[2]
&$\omega_{p}$&$\lambda_{p}$ &$\omega_{p}$&$\lambda_{p}$
&$\omega_{p}$&$\lambda_{p}$

\\\cline{1-2}\cline{3-4}\cline{5-6}\cline{7-8}

87.1(1)B$_{1g}$&-&26.3(2)&0&22.5(2)&0.027
&10.0(2)&0.047\\

74.5(2)E$_{2g}$&0.907 &27.0(2)&0.06 &28.4(2)&0.029
&16.8(2)&0.074\\

49.8(1)A$_{2u}$&- &39.0(1)&0.08&39.2(1)&0.109
&21.0(2)&0\\

40.7(1)E$_{1u}$&- &39.7(2)E$_{1u}$&0&39.3(2)&0.010
&25.0(2)&0.003\\

&&47.6(1)&0 &48.9(2)&0.026
&33.6(1)&0\\

&&49.3(1)A$_{2u}$&0&51.0(1)&0.019
&36.5(1)&0.124\\

&&68.0(2)&0.51&53.0(1)&0.005
&41.2(2)&0.002\\

&&73.4(2)E$_{2g}$&0.46&72.1(2)&0.506
&55.4(1)&0\\

&&79.2(1)&0.03&90.4(1)&0.002
&83.1(1)&0\\

&&86.3(1)B$_{1g}$&0&161.8(1)&0.088
5&89.121)&0.153\\

&&&&&&105.3(1)&0.131\\

&&&&&&215.8(1)&0\\\hline

\multicolumn{2}{c}{$\lambda=0.907$}
&\multicolumn{2}{c}{$\lambda=1.14$}
&\multicolumn{2}{c}{$\lambda=0.821$}
&\multicolumn{2}{c}{$\lambda=0.534$}

\\\hline

\multicolumn{2}{c}{$\omega_{log}=74.5$}
&\multicolumn{2}{c}{$\omega_{log}=64.5$}
&\multicolumn{2}{c}{$\omega_{log}=65.6$}
&\multicolumn{2}{c}{$\omega_{log}=48.9$}

\\\hline

\multicolumn{2}{c}{$T_{C}=39.4$}
&\multicolumn{2}{c}{$T_{C}=40.73$}
&\multicolumn{2}{c}{$T_{C}=18.31$}
&\multicolumn{2}{c}{$T_{C}=1.62$}

\end{tabular}
\end{ruledtabular}
\end{table*}

The relaxed geometries for structures with partial substitution of
boron atoms with one and two oxygen atoms, {Mg$_{2}$B$_{3}$O} and
{Mg$_{2}$B$_{3}$O$_{2}$}, as well as {Mg$_{2}$B$_{4}$}, are shown
in Fig. 2. Other oxygen concentrations were tried as well (i.e.
{Mg$_{2}$B$_{3}$O$_{3}$}, where one additional oxygen was inserted
in the boron plane), but these resulted in a significant
distortion of the {MgB$_{2}$}-like symmetry and did not reproduce
features observed in the coherent oxygen precipitates. We found
that the incorporation of oxygen changes the well-defined B-B
planes by the formation of {BO$_{x}$} units in the
{Mg$_{2}$B$_{3}$O$_{x}$} structures. Both {Mg$_{2}$B$_{3}$O$_{x}$}
structures belong to the space group P3m1 ({\em{C}$_{3v}^{1}$},
No. 156). The relevant structural parameters for the three
structures considered are shown in Table \ref{table1}.
Introduction of oxygen considerably increases the distance between
neighboring Mg planes, but decreases the distance between the
other Mg planes, which are above and below the boron layer left
intact.  As a result, the lattice stretches in the c-direction
compared to {Mg$_{2}$B$_{4}$}. Such behavior was reported by Eom
{\em{et. al.}} where thin films rich in oxygen showed a higher c
lattice parameter suggesting that boron can be substituted by
oxygen \cite{eom2001-01}. The calculated ratios of the distances
between the Mg-{BO$_{x}$}-Mg and Mg-BB-Mg layers are 1.22 and 1.49
for the theoretical {Mg$_{2}$B$_{3}$O} and
{Mg$_{2}$B$_{3}$O$_{2}$} structures, respectively, while the
estimate from the experimental image is 1.14 $\pm$ 0.05. The
overestimated theoretical values for the ratios compared with the
experimental can be explained in two different ways: (i) pressure
effects from the surrounding bulk into the precipitate, which
shrinks the lattice parameter in the c-direction, and (ii) the
concentration of oxygen in the {BO$_{x}$} planes is slightly lower
than the ones calculated here, where x $<$ 1.

We also investigated the phase stability of the
{Mg$_{2}$B$_{3}$O$_{x}$} compounds relative to atomic constituents
and bulk {MgB$_{2}$}. Both {Mg$_{2}$B$_{3}$O$_{x}$} structures are
stable against phase separation into their atomic constitutes.
However, the magnitudes of the binding energies per atom for
{Mg$_{2}$B$_{3}$O} are found to be 0.16 eV and 0.63 eV smaller
compared to that of {MgB$_{2}$} calculated using atomic oxygen,
and molecular oxygen, respectively. On the other hand, the
magnitudes of the binding energies per atom for
{Mg$_{2}$B$_{3}$O$_{2}$} are 0.11 eV higher, and 0.71 eV lower
compared to that of {MgB$_{2}$} calculated using atomic oxygen,
and molecular oxygen, respectively. These calculations suggest
that (i) the theoretical {Mg$_{2}$B$_{3}$O$_{x}$} structures are
not stable against phase separation into {MgB$_{2}$} and molecular
oxygen, and (ii) {Mg$_{2}$B$_{3}$O$_{2}$} is stable with respect
to phase separation into {MgB$_{2}$} and atomic oxygen. This does
not, however, mean that these structures have to be ruled out as
possible candidates for coherent oxygen precipitates, since they
can still exist as metastable phases in bulk {MgB$_{2}$}.

\begin{figure}
\includegraphics[scale=0.5]{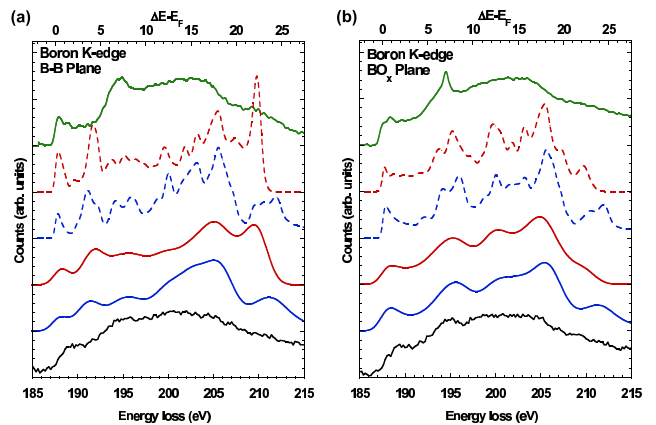}
\caption{(color). Experimental Boron K-edge spectra of coherent
oxygen precipitate (black) compared with calculated density of
states of {Mg$_{2}$B$_{3}$O} (blue)  and {Mg$_{2}$B$_{3}$O$_{2}$}
(red) (a) B-B plane, and (b) {BO$_{x}$} plane. Solid (dashed)
lines have an energy resolution of 1.2 (0.4) eV. Experimental
monochromated spectra (green) with an energy resolution of 0.3 eV
are shown for comparison for (a) bulk MgB$_{2}$, and (b) a
precipitate region with oxygen. Notice that the pre-peak in the
Boron K-edge is completely resolved for (a) and (b). Changes due
to oxygen in the monocromated spectra (b) are clearly notorious,
change of intensity in the pre-peak and the peak at 195 eV.}
\end{figure}

Figure 3 shows the calculated unoccupied boron DOS for the
theoretical {Mg$_{2}$B$_{3}$O$_{x}$} nanostructures compared with
the experimental boron K-edge spectra taken from the B-B plane
(darker columns in Fig. 1) and {BO$_{x}$} plane (brighter columns
in Fig. 1) respectively. Each experimental spectrum is the sum of
seven individual spectra with an acquisition time of 1 sec. and an
energy resolution of 1.2 eV. The spectra have been summed up to
increase the signal to noise ratio, and background subtracted and
corrected for multiple-scattering contributions
\cite{klie2002-01}. To obtain a better comparison between theory
and experiment, we have calculated the boron $p_{xy}$ and $p_{z}$
states. We then took into account the scattering momentum transfer
dependence of the electron microscope optical conditions and
crystal orientation to weigh the boron {p$_{xy}$} and {p$_{z}$}
states according to the experimental setup \cite{klie2003-01}.
Analysis of the spectra for energies in the range of 15-20 eV
above the Fermi level was not performed, since multiple scattering
and lifetime broadening limit the resolution of the spectra, which
makes comparison between experiment and theory uncertain in that
regime \cite{klie2003-01}. The theoretical spectra was obtained in
the $Z$ approximation, which has worked well for bulk MgB$_2$
\cite{klie2003-01}, have been convoluted for an energy resolution
of 1.2 eV and 0.4 eV (shown for comparison purposes only). Figure
3(a) shows the experimental and theoretical spectra for the B-B
plane, where both structures, {Mg$_{2}$B$_{3}$O} and
{Mg$_{2}$B$_{3}$O$_{2}$} (energy resolution of 1.2 eV), have good
agreement with experiment. The pre-peak of the boron K-edge at 189
eV in the experiment is well reproduced in both theoretical
structures. A peak at 192 eV, which is present in both theoretical
spectra, is not resolved in the experimental spectra. This is
possibly due to the low signal to noise ratio of the experimental
spectra. Figure 3(b) shows the spectra for the {BO$_{x}$} plane,
where the enhancement of the peak at 195 eV, which was previously
attributed to oxygen \cite{klie2002-01,liao2003-01,zhu2002-01}, is
very well predicted by both theoretical structures. This peak has
a lower intensity for the B-B plane spectra (Fig. 3(a)). Finally,
the pre-peak in the {BO$_{x}$} plane experimental spectra shows an
enhancement in intensity compared with the B-B plane experimental
spectra. Here, both theoretical spectra present a remarkable
agreement with the experiment, especially the {Mg$_{2}$B$_{3}$O}
structure where the difference in intensity of the pre-peak
between the B-B plane and the {BO$_{x}$} plane spectra is more
pronounced. Additionally, monochromated spectra (energy resolution
of 0.3 eV) from bulk {MgB$_{2}$} and a region containing oxygen
are shown in Fig. 3(a) and Fig. 3(b), respectively. Notice that
the pre-peak is completely resolved for Fig. 3(a) and Fig. 3(b).
Furthermore, the monochromated spectrum (Fig. 3(b)) presents an
enhancement due to oxygen of the peak at 195 eV.

To investigate the effect of oxygen incorporation by partial boron
substitution on the superconducting properties of {MgB$_{2}$}, we
first calculated the phonon modes and electron phonon coupling
parameters for {Mg$_{2}$B$_{4}$}, {Mg$_{2}$B$_{3}$O}, and
{Mg$_{2}$B$_{3}$O$_{2}$}. The results are shown in Table
\ref{table1}. In order to obtain the transition temperatures
{\em{T$_{C}$}} of the theoretical {Mg$_{2}$B$_{3}$O$_{x}$}
nanostructures we used the McMillian expression
\cite{mcmill1968-01}
\begin{equation}
T_{C}=\frac{\omega_{log}}{1.2}\exp{\left(\frac{-1.04(1+\lambda)}{\lambda-\mu^{*}(1+0.62\mu^{*}\lambda)}\right)},
\end{equation}
where $\omega_{log}$ is the logarithmic average frequency,
{$\lambda$} is the total electron phonon coupling defined as
$\lambda=\sum\limits_{p\in\Gamma}\lambda_{p}$, with
{$\lambda_{p}$} being the electron phonon coupling for each mode
at $\Gamma$, and {$\mu^{*}$} represents the Coulomb repulsion
interaction. We used the value of {$\mu^{*} = 0.19$}. In the case
of {MgB$_{2}$}, the Eliashberg function has a sharp peak around
the {\em{E$_{2g}$}} mode \cite{choi2003-01}; therefore, it is
sufficient to take the {\em{E$_{2g}$}} mode energy as
{$\omega_{log}$} to obtain the transition temperature
{\em{T$_{C}$}} \cite{an2001-01,yildi2001-01}. In fact, using only
this mode, not only {\em{T$_{C}$}} but also the isotope effect and
pressure dependence of {\em{T$_{C}$}} can be successfully
explained \cite{yildi2002-01}. However, for the
{Mg$_{2}$B$_{3}$O$_{x}$} nanostructures, the {\em{E$_{2g}$}} mode
is no longer well defined and {$\omega_{log}$} has to be averaged
throughout all the phonon modes. We approximated {$\omega_{log}$}
as
{$\omega_{log}\approx\exp{\left(\frac{1}{\lambda}\sum\limits_{n}\ln\left(\omega_{p}\right)\lambda_{p}\right)}$},
and only calculated the phonons at the zone center {$\Gamma$}. We
obtained transition temperatures {\em{T$_{C}$}} of 40.73 K, 18.31
K, and 1.62 K for {Mg$_{2}$B$_{4}$}, {Mg$_{2}$B$_{3}$O}, and
{Mg$_{2}$B$_{3}$O$_{2}$}, respectively.

In conclusion, we have proposed two different structures,
{Mg$_{2}$B$_{3}$O} and {Mg$_{2}$B$_{3}$O$_{2}$} as possible
candidates for the {Mg$_{2}$B$_{3}$O$_{x}$} nanostructures
observed experimentally in bulk {MgB$_{2}$} \cite{klie2002-01}.
Both structures present an increase of the c lattice parameter
with respect to {Mg$_{2}$B$_{4}$}, where the {Mg$_{2}$B$_{3}$O}
structure is closer to the experimental value. The experimental
spectra for the B-B plane and {BO$_{x}$} plane are well explained
by the {Mg$_{2}$B$_{3}$O} and {Mg$_{2}$B$_{3}$O$_{2}$} structures.
Both structures are good candidates for coherent oxygen
precipitates, although their superconductivity properties differ.
{Mg$_{2}$B$_{3}$O$_{2}$} precipitates would behave as a pinning
center for temperatures higher than 2 K, while {Mg$_{2}$B$_{3}$O}
precipitates would start to behave as a pinning center for
temperatures higher than 18 K. Our results show that oxygen
impurities are likely to be present everywhere in {MgB$_{2}$},
highlighting the importance of oxygen as an agent in the creation
of flux pinning centers and as a factor of improvement upon the
critical current density {\em{J$_{C}$}}.

J. C. Idrobo thanks H. J. Choi for useful
discussions and R. Erni for assistance on the
FEI G2 Tecnai
electron microscope. This work was supported by DOE, NSF, and LBNL.


\begin{thebibliography}{99}
\bibitem{klie2002-01} R. F. Klie, J. C. Idrobo, N. D. Browning, A. Serquis, Y. T. Zhu,
X. Z. Liao, and F. M. Mueller, Appl. Phys. Lett. {\bf80}, 3970
(2002).

\bibitem{naga2001-01}J. Nagamasu, N. Nakagawa, T. Muranaka, Y. Zenitani, and J.
Akimitsu, Nature {\bf410}, 63 (2001).

\bibitem{wang2002-01}J. Wang, Y. Bugoslavsky, B. Berenov, L. Cowey, A. D. Caplin, L. F.
Cohen, L. D. Cooley, X. Song, and D. C. Larbalestier, Appl. Phys.
Lett. {\bf81}, 2026 (2002).

\bibitem{feng2001-01}Y. Feng {\em{et. al.}}, Appl.Phys. Lett. {\bf79}, 3983
(2001).

\bibitem{take2001-01}T. Takenobu, T. Ito, D. H. Chi, et. al., Phys. Rev. B {\bf64}, 134513
(2001).

\bibitem{slusky2001-01}J.S Slusky, {\em{et. al.}} Nature {\bf410}, 343
(2001).

\bibitem{solta2002-01}S. Soltanian, X. L. Wang, J. Hovart, {\em{et. al.}} Physica C {\bf382}, 187
(2002).

\bibitem{yan2003-01}Y. Yan and M. M. Al-Jassim., Phys.  Rev. B {\bf67}, 212503
(2003).

\bibitem{sharma2002-01}P. A. Sharma, N. Hur, Y. Horibe, C. H. Chen, B. G. Kim, S. Guha,
M. Z. Cieplak, and S-W. Cheong, Phys. Rev. Lett. {\bf89}, 167003
(2002).

\bibitem{klie2001-01}R. F. Klie, J. C. Idrobo, N. D. Browning, K. A. Regan, N. S.
Rogado, and R. J. Cava, Appl. Phys. Lett. {\bf79}, 1837 (2001).

\bibitem{liao2003-01}X. Z. Liao, A. Serquis, Y. T. Zhu, J. Y. Huang, L. Civale, D. E.
Peterson, F. M. Mueller, and H. F. Xu, Journal of Appl. Phys.
{\bf93}, 6208 (2003).

\bibitem{larba2001-01}D. Larbalestier, A. Gurevich, M. Feldmann, and A. Polyanskii,
Nature {\bf414}, 368 (2001).

\bibitem{canfi2001-01}P. C. Canfield, D. K. Finnemore, S. L. Bud'ko, J. E. Ostenson, G.
Lapertot, C. E. Cunningham, and C. Petrovic, Phys. Rev. Lett.
{\bf86}, 2423 (2001).

\bibitem{grasso2001-01}Grasso {\em{et. al.}} Appl. Phys. Lett. {\bf79}, 230
(2001).

\bibitem{mija2002-01}D. Mijatovic, A. Brinkman, I. Oomen, G. Rijnders,H. Hilgenkamp, H.
Rogalla, and D. H. A. Blank, Appl. Phys. Lett. {\bf80}, 2141
(2002).

\bibitem{eom2001-01}C. B. Eom, M. K. Lee, J. H. Choi, L. J. Belenky, X. Song, L. D.
Cooley, M. T. Naus, S. Patnaik, J. Jiang, M. Rikel, A. Polyanskii,
A. Gurevich, X. Y. Cai, S. D. Bu, S. E. Babcock, E. E. Hellstrom,
D. C. Larbalestier, N. Rogado, K. A. Regan, M. A. Hayward, T. He,
J. S. Slusky, K. Inumaru, M. K. Haas, and R. J. Cava, Nature
{\bf411}, 588 (2001).

\bibitem{browning1993-01}For further information in the experimental techniques see N. D.
Browning, M. F. Chisholm, and S. J. Pennycook, Nature {\bf366},
143 (1993).

\bibitem{kresse1996-01}G. Kresse and J. Furthmuller, Phys. Rev. B {\bf54}, 11169
(1996).

\bibitem{perdew1992-01}J. P. Perdew, J. A. Chevary, S. H. Vosko, K. A. Jackson, M. R.
Pederson, D. J. Singh, and C. Fiolhais, Phys. Rev. B {\bf46}, 6671
(1992).

\bibitem{baroni2001-01}S. Baroni, S. Gironcoli, A. Dal Corso, and
P. Giannozzi, Rev. Mod. Phys. {\bf73}, 515 (2001).

\bibitem{klie2003-01}R. F. Klie, H. Su, Y. Zhu, J. Davenport, J. C. Idrobo, N. D.
Browning, and P. D. Nellist, Phys. Rev. B {\bf67}, 144508 (2003).

\bibitem{zhu2002-01}Y. Zhu, A. R. Moodenbaugh, G. Schneider, J. W. Davenport, T. Vogt, Q. Li, G. Gu,
D. A. Fisher, J. Tafto, Phys. Rev. Lett. {\bf88}, 247002 (2002).

\bibitem{mcmill1968-01}W. L. McMillan, Phys. Rev. {\bf167}, 331
(1968).

\bibitem{choi2003-01} H. J. Choi, D. Roundy, H. Sun, M. L. Cohen, and S. G. Louie,
Nature {\bf 418}, 758 (2002); J. Choi, M. L. Cohen, and S. G. Louie
Physica C {\bf385}, 66 (2003).

\bibitem{an2001-01}J. M. An and W. E. Pickett, Phys. Rev. Lett.
{\bf86}, 4366 (2001).

\bibitem{yildi2001-01}T. Yildirim, O. Gulseren, J. W. Lynn, C. M. Brown, T. J. Udovic,
Q. huang, N. Rogado, K. A. Regan, M. A. Hayward, J. S. Slusky, T.
He, M. K. Haas, P. Khalifah, K. Inamaru, and R. J. Cava, Phys.
Rev. Lett. {\bf87}, 037001 (2001).

\bibitem{yildi2002-01}T. Yildirim and O. Gulseren, J. Phys. Chem.
Solids {\bf63}, 2201 (2002)

\end{thebibliography}
\end{document}